\documentstyle[aaspp4]{article}
\begin{document}

\title{The Astronomer's Telegram: A Web-based Short-Notice Publication
 System for the Professional Astronomical Community}
\author{Robert E. Rutledge}
\affil{Department of Astronomy,  University of California, Berkeley,
CA, 94720 \\
  rutledge@astron.berkeley.edu}

\begin{abstract}
{\it The Astronomer's Telegram} (ATEL; http://fire.berkeley.edu:8080/)
is a web based short-notice ($<$4000 characters) publication system
for reporting and commenting on new astronomical observations,
offering for the first time in astronomy {\em effectively
instantaneous distribution of time-critical information for the entire
professional community}.  It is designed to take advantage of the
World Wide Web's simple user interface and the ability of computer
programs to provide nearly all the necessary functions.  This makes
ATEL fast, efficient, and free.  In practice, one may post a Telegram,
which is instantly ($<$1 second) available at the Web-site, and is
distributed by email within 24 hours through the Daily Email Digest,
which is tailored to the subject selections of each reader.  In
addition, authors reporting new outbursts of transients or coordinates
of new objects (for example, gamma-ray bursts or microlensing events)
may request distribution by Instant Email Notices, which instantly
($\sim$ minutes) distributes their new Telegram by email to
self-identified workers interested in the same topic.  This speed in
distribution is obtained because no editing or reviewing is performed
after posting -- the last person to review the text before
distribution is the author.  Telegrams are enumerated chronologically,
permanently archived, and referenceable.  While ATEL will be of
particular use to observers of transient objects (such as gamma-ray
bursts, microlenses, supernovae, novae, or X-ray transients) or in
fields which are rapidly evolving observationally, there are no
restrictions on subject matter.

\end{abstract}


\section{Introduction}

In recent years, use of the World Wide Web (WWW) among astronomers has
become commonplace.  Most individuals have their own web site, where
they maintain a library of their recent preprints and professional
information.  Data from observatories is often distributed through a
web site, as well as observing application forms; many observatories
now require submission of these application forms through a web
interface.  Nearly all major astronomical journals use the web to
distribute their refereed articles, often before they are available in
paper form. These include {\it The Astrophysical Journal}, the main
journal, Supplement and Letters
(http://www.journals.uchicago.edu/ApJ/journal/); {\it Astronomy \&
Astrophysics}
(http://link.springer.de/link/service/journals/00230/index.htm); {\it
The Astronomical Journal}
(http://www.journals.uchicago.edu/AJ/journal/); {\it Publications of
the Astronomical Society of the Pacific}
(http://www.journals.uchicago.edu/PASP/journal/index.html), and {\it
Science} (http://www.sciencemag.org/); although not (yet) {\it Monthly
Notices of the Royal Astronomical Society}, or {\it Publications of the
Astronomical Society of Japan}. The journal {\it Nature}
(http://www.nature.com/) permits
reading of its abstracts, but not text, online simultaneously with the
print publication.

A web-based electronic journal of refereed articles - {\it New
Astronomy} -- is distributed by the WWW as well as in paper form, but
departs from traditional paper-based journals largely by permitting
greater flexibility in the medium of presentation (color figures,
time-evolved video, sound, and other media  which can be distributed
by the WWW).

There also exists a unique Internet resource which permits the
widespread distribution of pre-prints from a single site -- known as
the e-Print archive (http://xxx.lanl.gov/).  This allows astronomers,
and practicing scientists in other fields, to distribute their
pre-printed articles prior to publication, and often prior to a
reading by a referee. This site has become the {\it de facto} means of
communicating new results to the astronomical community, because those
wishing to remain at the forefront of their field require the most
up-to-date work as soon as it is available and are competent to
evaluate the quality of the presented work themselves.

The main advantage of the WWW and Internet distribution over the more
traditional means of article publication has not yet been
systematically exploited.  This advantage is {\em effectively
instantaneous distribution of time-critical information}.  This
advantage is most useful to observational astronomers who study
transient objects -- such as gamma-ray bursts, microlensing events,
galactic X-ray sources, novae, or super-novae -- and to fields which
are rapidly evolving observationally, such as (presently) brown
dwarfs, extra-solar planets, and milli-second pulsars.  This advantage
is required more for the distribution of short observational reports
with a minimal amount theoretical interpretaion than for the detailed
presentation of results and intricate (and perhaps controversial)
interpretation, for which refereed articles in traditional print
journals are useful.

\subsection{Previous Distribution of Time Critical Information}

Time critical information -- such as the coordinates of new
discoveries, recent intensity measurements in a variety of wave bands,
and rapid theoretical interpretation of behavior which has immediate
implications for observations -- has hitherto been distributed through
formal and informal networks.

The most widely used formal network is the electronic International
Astronomical Union Circulars (IAUCs;
http://cfa-www.harvard.edu/cfa/ps/cbat.html).  Submitted text is
reviewed and edited by the editors, sometimes returned to the author
for changes, and distributed electronically between a few hours and a
few days after submission.  For some types of object ({\it e.g.}
gamma-ray bursts, microlensing events), a few days delay time is the
difference between a well studied object and a missed observational
opportunity.  The IAUCs also operate as a referenceable repository of
information, which makes the author accountable for the presented
results, providing motivation for the author to present reliable
information.

Most fields of transient objects have developed their own informal networks
-- sometimes consisting of email exploders, which redistribute
messages to a list of interested individuals, or even as informal as a
single individual who emails received messages to everyone they know.
These networks are typically advertised by ``word of mouth'', and are
not often widely familiar, limiting the community of workers, and thus
the coverage which one can obtain.  Also, such informal networks are
susceptible to simple breakdowns in communication -- due to absences
of key distributers, and the difficulty in manually maintaining a
relevant email list over the long term.  Also, due to the informality
of the information distribution, the distributed information is not
static, archived, or referenceable; often, one does not know who
produced the coordinates at which one is to re-direct one's telescope
for follow-on observations.  This limits the reliability of and
subsequent confidence in this information, which is essential to other
observers evaluating the information for the possibility of further
observations.

\section{The Astronomer's Telegram}

ATEL is a means to make use of the advantages offered by automation
and the WWW for the information distribution needed by astronomers.

It is designed with the problem in mind that the limiting factor in
observations of transient objects is the time-scale of distribution of
observational information -- discoveries, measurements, evolution,
availability of finding charts, predictions for future behavior, and
rapid pertinent interpretation of the observations which immediately
affect observational planning.  This time-scale is loosely defined as
the elapsed time between when the authors realize when they have
information which will be useful to other observers, and when these
other observers receive that information.

ATEL eliminates what has hitherto been the determining factor in this
time-scale -- time spent between the submission of the information and
its distribution -- by eliminating all people between the author and
the means of distribution (the WWW and email).

Thus, the design goals of ATEL are to: 

\begin{itemize}
\item maximize the reliability and relevance of posted information;
this is done by restricting the ability to post to professional
astronomers, requiring registration prior to posting of a username and
password or PGP public key, with the owner's identity and professional
status verified prior to activation.  Either an activated username and
passsword or a signature from an activated PGP key is required when a
Telegram is posted. 
\item provide a long-term, stable means of distribution. 
\item minimize the distribution time and maximize availability of
posted Telegrams; this is accomplished by having all functions
performed by computer programs, executed through the Web by distant users. 
The delay time for publication is therefore set by the computational
speed of the host computer and the speed of the WWW protocols, and
presently is $<$ 1 second.
\item provide an easy means of posting, reading and searching Telegrams; 
\item performing all these functions at zero cost at the point of
use to both authors and readers. 
\end{itemize}

\subsection{Functions and Use}

All interactions with ATEL occur through the WWW (at
http://fire.berkeley.edu:8080/) where one may:

\begin{itemize}
\item Read posted Telegrams, sortable by subject, date, author, or
user-defined keywords. 
\item Sign up for the Daily Email Digest; by specifying those subjects
of interest, ATEL will email to you, each day,  those Telegrams
received during the previous 24-hour period.  No email is sent if no such
Telegrams are received. 
\item Sign up for the Instant Email Notices; information which is
extremely time critical, such as the coordinates of a newly discovered
object, are emailed at author request to those who have asked to
receive these in selected subject areas. 
\item Post a Telegram for instant reading through the Web and/or Instant
Email Notices, and distributed within 24 hours in the Daily Email
Digest.   Posted Telegrams may make use of HTML formatting, including
links to other sites, where supplementary information might be
provided.  Authors should keep in mind that many readers will see
their information in ASCII email, where some HTML formatting (such as
TABLE) may not be clear.  Several HTML commands -- (specifically IMG,
META, APPLET, FRAME and HR) are presently not permitted, to insure the
static nature of the content, and to keep the Web page layout uniform.
\item Submit for registration a username and password or PGP key, with which
(once the identity and status of the user is verified)
the user may post Telegrams. 
\end{itemize}

Telegrams are posted and instantly available, which is possible only
because no review of the material (i.e.  editing or refereeing) takes
place after submission.  The version which the author posts is the
version which is available at the web site, and no modifications by
anyone may be made after the posting (HTML commands, however, are
presently stripped from a Telegram prior to emailing the Daily Email
Digest or the Instant Email Notices).  Authors who discover errors
after posting must post an additional Telegram, delineating the error.

\section{Conclusions}

{\em The Astronomer's Telegram} offers a publication method which was
not previously available to the astronomical community with
effectively instantaneous distribution.  This is useful, most notably
to observers of transient objects and to fields which are rapidly
evolving.  There are no restrictions on subject matter, and so it may
be used by all astronomical professionals.

\acknowledgements

I am grateful for useful comments on this paper by G. Magnier, and
C. Moore.  I am indebted to D. Fox, J. Kommers, C. Moore, and
P. Wojdowski, for advice on the design and implementation of the web
site, and to L. Bildsten for providing a location for the web site.

\end{document}